\RequirePackage{fix-cm}
\documentclass[]{svjour3}                     % onecolumn (standard format)
\smartqed  % flush right qed marks, e.g. at end of proof
\usepackage{pdflscape}
\usepackage{fixltx2e}
\usepackage{xcolor}
\usepackage{hyperref}
\usepackage{graphicx}
\usepackage{amsmath}
\usepackage{multirow}
\usepackage{tabularx}
\usepackage{amsfonts}
\usepackage{amssymb}
\usepackage{multirow}
\usepackage{epstopdf}
\usepackage[caption=false]{subfig}
\usepackage{float}
\usepackage[utf8]{inputenc}
\usepackage{caption}
\usepackage{algorithm}
\usepackage{ amssymb }
\usepackage{algpseudocode}
\usepackage[numbers, sort]{natbib}
\usepackage{tikz}

\journalname{Journal of Latex Class}
\begin{document}
	
	\title{The Impact of Copycat Attack on RPL based 6LoWPAN Networks in Internet of Things}
	
	\titlerunning{Impact of Copycat Attack on RPL protocol}        % if too long for running head
	
	\author{Abhishek Verma$^{1,2,*}$\and
		Virender Ranga$^{2}$
	}

	\institute{	$^{1}$Department of Computer Engineering and Applications, Institute of Engineering \& Technology, GLA University, Mathura, India \\
	$^{2}$Department of Computer Engineering, National Institute of Technology Kurukshetra, India \\
	\email{$^{*}$abhiverma866@gmail.com} \\ 
	\email{virender.ranga@nitkkr.ac.in} \\         %  \\
	}
	
	\date{Received: xx-xx-xxxx / Accepted: xx-xx-xxxx}
	% The correct dates will be entered by the editor
	\maketitle
\begin{abstract}
	IPv6 Routing Protocol for Low-Power and Lossy Networks (RPL) is the standard network layer protocol for achieving efficient routing in IPv6 over Low-Power Wireless Personal Area Networks (6LoWPAN). Resource-constrained and non-tamper resistant nature of smart sensor nodes makes RPL protocol susceptible to different threats.  An attacker may use insider or outsider attack strategy to perform Denial-of-Service (DoS) attacks against RPL based networks. Security and Privacy risks associated with RPL protocol may limit its global adoption and worldwide acceptance. A proper investigation of RPL specific attacks and their impacts on an underlying network needs to be done. In this paper, we present and investigate one of the catastrophic attacks named as a copycat attack, a type of replay based DoS attack against the RPL protocol. An in-depth experimental study for analyzing the impacts of the copycat attack on RPL has been done. The experimental results show that the copycat attack can significantly degrade network performance in terms of packet delivery ratio, average end-to-end delay, and average power consumption. To the best of our knowledge, this is the first paper that extensively studies the impact of RPL specific replay mechanism based DoS attack on 6LoWPAN networks.\footnote{The final publication is available at https://link.springer.com/article/10.1007/s00607-020-00862-1} 
\end{abstract}


\section{Introduction}\label{Introduction}
Drastic growth in the development of Internet-based technologies has been observed in recent decades. With the start of the era when huge sized computers required too much of human intervention to the era where small-sized smart devices that operate without any human intervention show that there has been a significant development in computer and Internet-based technologies. In recent years, a new networking paradigm termed as the Internet of Things (IoT) \cite{VermaIEEE, ashton2009internet, IoTCore1, miloslavskaya2019internet} has evolved. IoT is currently seen as a fast-evolving networking paradigm that consists of smart devices that communicate to control the physical surroundings. \textcolor{black}{Ericcson is expecting $ 3.5$ Billion cellular IoT connections by $ 2023 $ and Global Data forecasts the global market for IoT technology to reach \$$ 318 $ Billion by $ 2023 $ \cite{GlobalData}}. \textcolor{black}{A broad range of applications is being developed worldwide to make human life safer and comfortable, e.g., e-health, smart home, smart grid, and smart city.} \textcolor{black}{Such applications are changing human lives and making it easier to live. A drastic increase in the number of smart devices has lead to severe cybersecurity risks for smart applications.} These risks expose users' security and privacy to attackers \cite{, VermaIEEE, Raoof, Alaba2017, AIREHROUR2016198, ziegeldorf2014privacy, Yang7902207, qadri2020limitations, miloslavskaya2019internet}. Moreover, when healthcare and smart grid applications are exposed to attackers, such situations may even lead to life-endangering cases for human beings \cite{6305831}.

\textcolor{black}{Most of the IoT applications are enabled by a large scale deployment of IPv6 over Low-Power Wireless Personal Area Networks (6LoWPAN), i.e., a type of Low Power and Lossy Networks (LLNs).} LLNs have lossy and low throughput communication links. These networks are enabled by resource constrained devices \cite{COLAKOVIC2018, winter2012rpl}. The resource constrained devices (nodes) operate on low power, require less energy, have small on-board memory, and low computational capabilities \cite{Musaddiq}. The characteristics of LLNs make traditional routing protocols, including Adhoc On-Demand Distance Vector (AODV), Dynamic Source Routing (DSR), Open Shortest Path First (OSPF) are unsuitable for LLNs \cite{tripathi2014design}. The IPv6 Routing Protocol for Low-Power and Lossy Networks (RPL) was standardized (RFC $ 6550 $) in 2012 \cite{winter2012rpl}, and it enables energy efficient routing in LLNs. However, the characteristics like self-organization, self-healing, open nature, and resource constrained nature expose RPL to a different type of routing attacks, which may compromise users' security and privacy \cite{RIAHISFAR2018118, Verma2019, shafique2018detection,sahay2020novel}. In recent years, vulnerabilities and threats associated with RPL have been rigorously explored by the researchers. In this paper, the main focus is on packet replay attacks, which may affect the Quality of Service (QoS) of real-time wireless networks.

In wireless network based replay attacks \cite{kannhavong2007survey,4599222}, an attacker node eavesdrops on broadcast messages of legitimate nodes and later sends the previously eavesdropped message to its neighbor nodes. Thus, the victim nodes are forced to believe that the information they received is fresh, which is not. This situation may lead to undesirable consequences like packet loss and degraded network performance. In RPL, the replay attack is mainly performed by replaying control messages rather than data messages. Thus, the victim nodes are forced to update their routing tables with obsolete routing information that leads to the creation of non-optimal topology and degraded routing performance.

\textcolor{black}{In this paper, an RPL protocol specific replay attack named as the copycat attack is presented, and its impact on the underlying network is analyzed.} The copycat attack is a Denial-of-Service (DoS) attack, which can catastrophically degrade the performance of RPL based networks in terms of Packet Delivery Ratio (PDR), Average End-to-End Delay (AE2ED), and Average Power Consumption (APC). To launch this attack, an attacker node eavesdrops DODAG Information Object (DIO) control messages of legitimates nodes, and later sends the previously eavesdropped DIO messages many times with fixed replay interval. In this manner, an attacker induces victim nodes to choose non-optimal parents, which consequently leads to the creation of non-optimal routes.  Also, an attacker does not need to have any high range radio antenna or any other specialized hardware to perform copycat attacks. \textcolor{black}{The novelty of this paper is that it presents a novel routing attack that targets RPL protocol and analyzes the attack impact in terms of prominent performance evaluation metrics. In addition, this paper also suggests some possible solutions that can be used to mitigate the proposed routing attack.}    

The remaining paper is structured in the following way. Section \ref{background} presents the background of this study. In Section \ref{Related Work}, some relevant research works are described. The copycat attack is presented in Section \ref{Copycat Attack}. A detailed discussion of the experimental evaluation of the copycat attack is presented in Section \ref{Experimental Evalutaion}. Some possible solutions for addressing copycat attacks are discussed in Section \ref{PossibleSol}. The paper is concluded in Section \ref{Conclusions and Future Scope}.

\subsection{Motivation}

There are several works that have suggested replay attacks with respect to RPL protocol \cite{Le2013, Mayzaud2016Taxanomy, Raoof}. However, none of the works has performed an in-depth study on how such attacks take place in the RPL based networks and not analyzed the impacts of such attacks. This motivated us to pursue the current research work in order to fulfill this research gap.   

\subsection{Major Contributions}
The major contributions of the paper are mentioned below:

\begin{enumerate}
	\item An RPL specific replay mechanism based DoS attack named as the copycat attack is presented.
	\item An in-depth experimental study for analyzing the impacts of the copycat attack on RPL based 6LoWPAN networks is carried out. 
	\item Possible solutions to address copycat attacks have been suggested to motivate future research towards the development of defense solutions. 
\end{enumerate}

\section{\textcolor{black}{Background}}\label{background}
\textcolor{black}{This section provides brief overview of 6LoWPAN and RPL protocol.}

\subsection{\textcolor{black}{6LoWPAN}}
\textcolor{black}{6LoWPAN \cite{rfc6LoWPAN, olsson20146lowpan} combines IPv$ 6 $ and Low-Power Wireless Personal Area Network (LoWPAN). It allows resource constrained (power, memory, and processing) devices to communicate and share data wirelessly using IPv$ 6 $. It enables the smallest devices to be a part of the global IoT network.  Devices in 6LoWPAN can communicate with other $ 802.15.4 $ based devices, and the devices operating on the Internet connected through various wireless communication technologies like Wi-Fi, blacktooth, and  Near-field Communication \cite{hussain2017internet}. IPv$ 6 $ is used in IoT because of its large address space and its capability to provide global connectivity to nodes. However, it cannot be directly applied in IoT because of resource constrained devices. For this purpose, $ 6 $LoWPAN defines a compressed version of IPv$ 6 $ that is well suited for IoT. It specifies mechanisms for shortening IPv$ 6 $ $ 128 $-bit address, header compression to reduce transmission overhead, packet fragmentation for meeting the needs of IEEE $ 802.15.4 $ $ 127 $ byte Maximum Transmission Unit limit, and support for multi-hop packet delivery. $ 6 $LoWPAN acts as an adaptation layer in IoT communication protocol stack.}

\subsection{\textcolor{black}{RPL Routing Protocol}}\label{Overview of RPL}
\textcolor{black}{RPL is among popular routing protocols due to its flexible nature, QoS support, and energy-efficient routing capability \cite{Granjal, Palattella, Tomic, LLNMobility}. It has been standardized for 6LoWPAN \cite{Iova}, and its specifications are depicted in RFC $ 6550 $ \cite{winter2012rpl}. RPL is based on distance-vector and source routing protocols. It operates above IEEE $ 802.15.4 $ MAC layer protocol. RPL supports point-to-point, multipoint-to-point, and point-to-multipoint topologies. RPL virtually creates a  Destination Oriented Directed Acyclic Graph (DODAG) topological structure from nodes. DODAG is a loop-free and tree-like topological structure. A single IoT network contains multiple parallel \textit{RPLInstance} running at a single time, and a single \textit{RPLInstance} may contain multiple DODAGs. \textit{RPLInstance} is identified by \textit{RPLInstanceID} while DODAG is identified by DODAG ID which is a unique IPv6 address. The primary characteristics of the RPL are auto-configuration, self-healing, loop avoidance and detection, transparency, and support for multiple sinks. RPL uses four types of control messages for creating and maintaining DODAG:  (i) DIO, (ii) DODAG Information Solicitation (DIS), (iii) Destination Advertisement Object (DAO), and (iv) Destination Advertisement Object Acknowledgment (DAO-ACK). An Objective function (OF) is an important part of RPL, which is responsible for the selection and optimization of routes between DODAG nodes. OF utilizes different metrics and constraints for choosing optimal parent among various preferred choices, i.e., preferred parents. ETX Objective function (ETXOF) \cite{ETXOF}, Minimum Rank with Hysteresis Objective Function (MRHOF) \cite{MRHOF}, Objective Function Zero (OF0) \cite{OF0} are prominent OF of RPL. Each DODAG node is assigned a rank, which serves a significant role in DODAG management. Rank represents the node’s position with respect to the DODAG root. RPL specifies a strict rank rule. According to the rank rule, the rank value increases in DODAG's downward direction (root to leaves) and vice-versa. RPL's rank concept is used in RPL: (1) detection and resolving of routing loops; (2) to maintain a parent-child relationship; (3) to distinguish between parents and siblings; (4) to restore broken links. DIO message contains routing information that is needed by nodes to find existing \textit{RPLInstance} and RPL configuration parameters. DODAG node uses routing information contained in the DIO message to choose its preferred parent set. DIS message is used by a node to solicit a DIO message from an existing DODAG node. DIS message is primarily used when a new node wants to find nearby DODAG. DAO messages are used to create downward routes. DAO-ACK message is used by the root node to send an acknowledgment of the DAO message \cite{RPLNutshell}. RPL uses ``Trickle timer'' to limit the transmission of control messages in the network \cite{levis2011trickle} and minimize energy consumption. The trickle timer is reset in case of inconsistency detection, i.e., loops and link loss, change in parent set. The trickle timer interval is increased or decreased in case of a stable network and inconsistency detection, respectively. In the case of a stable network, the interval is increased in order to decrease the number of DIO's been transmitted in the network.  Whereas, upon detection of topological inconsistency, the interval is reduced to increase the number of DIO's to fix the inconsistency quickly \cite{IPSOAllianceRPL}.
}
\section{Related Work}\label{Related Work}

In recent years, there has been tremendous growth in literature, which focuses on IoT security perspectives, and the RPL protocol is among the most popular topics that have been studied. In \cite{Le2013}, a novel attack named neighbor attack has been presented. In a neighbor attack, an attacker node duplicates and multicast all the DIO messages it has received from its parent. In such a case, all the nearby nodes which receive the replayed DIO are forced to believe that the message is from some new neighbor. Further, if the replayed DIO contains favorable routing information like rank, then the victim node may add the out of range node as its preferred parent. Another variant of neighbor attack is proposed in \cite{Mayzaud2016Taxanomy} and termed as routing information replay attack. In this attack, an attacker node sends the obsolete DIO messages that contain old routing information. Thus, upon receiving any outdated DIO message, a victim node is induced to follow the stale and add un-optimized routes to its routing table \cite{Raoof}. As far as the literature is concerned, there are a limited number of works on the mitigation of replay based attacks. Perrey \textit{et al.}  \cite{landsmann2013topology} proposed a generic security scheme called Trust Anchor Interconnection Loop (TRAIL) for detecting and preventing topological inconsistency attacks (Version number, Rank spoofing, and Rank replay) in RPL based networks. TRAIL facilitates topology authentication in RPL. TRAIL enables each node to validate its upward routing path towards the root and detect any rank spoofing without relying on encryption chains. Le \textit{et al.} \cite{Le2016} proposed a specification based IDS for detecting Rank \cite{le2013impact}, Local repair, Neighbor, DIS and Sinkhole attacks. The proposed specification based Intrusion Detection System (IDS) consists of an Extended Finite State Machine generated from a semi-auto profiling technique. Tsao \textit{et al.} \cite{tsao2015security} suggested various attacks and their countermeasures specific to RPL protocol. However, no experimental study on the behavior of suggested attacks, and no performance evaluation of the suggested solutions is done in this study. Perazzo \textit{et al.} \cite{Perazzo2017} proposed a new type of attack against the RPL protocol named as DIO suppression attack. The idea behind this attack is to suppress the transmission of fresh DIO control messages required by the IoT nodes for exploring new optimized routing paths and the removal of stale paths. Verma \textit{et al.} analyzed the impact of DIS flooding attacks on RPL based networks. The authors also proposed a lightweight defense scheme named \textit{Secure-RPL} to defend 6LoWPAN network against DIS flooding attack \cite{verma2019addressing,verma2020mitigation}.   

%To the best of our knowledge, there is no work present in the literature that has performed extensive study on analyzing the impacts of replay attacks on RPL based networks.       

\section{Copycat Attack}\label{Copycat Attack}
The main goal of the copycat attack is to degrade the routing performance of RPL based networks so that the QoS of real-time applications is affected. To achieve this, an attacker may compromise a legitimate internal node and reprogram it to introduce an increased level of congestion and interference in the network. The attacker can also choose an outsider attack strategy to perform this attack. To launch a copycat attack, an attacker simply eavesdrops the DIO messages of nearby nodes, and later sends (multicast) the captured DIO message (with or without modification) many times with a fixed replay interval. The copycat attack can be of two types: 1) non-spoofed; 2) spoofed. In, ``non-spoofed copycat attack" the eavesdropped DIO is sent after modifying the source IP of the ICMPv6 packet containing the DIO message. The attacker sends the unmodified captured DIO with its own IP address in the ICMPv6 packet, which forces receiving (victim neighbors) nodes to believe that the packet is from a legitimate sender and makes them perform unnecessary routing related operations. Therefore, an attacker can drain the victim's resources and disrupt its normal packet forwarding behavior. The second type of copycat attack is termed as ``spoofed copycat attack". In this attack, the eavesdropped DIO is sent to neighbor nodes after replacing the source IP address of encapsulating IPv6 packet with the legitimate DIO sender's IP address, i.e., the sender of the eavesdropped DIO message. This makes the receiver believe that the sender of DIO is its in-range neighbor. The victim nodes may even try to add the out of range neighbor, assuming that it leads to an optimal route to the gateway. \textcolor{black}{In simple words, in non-spoofed copycat attack, the adversary uses its IP address as the source, and in spoofed copycat attack, the adversary uses the source IP address of a legitimate node as a source.} Both the attack types introduce heavy congestion and interference in their attack region, which consequently decreases PDR,  and increases AE2ED and APC of the underlying network. \textcolor{black}{The main difference between copycat attack and other replay attack variants (i.e., routing information replay and neighbor attack) lies in the frequency of replaying the packets and the packet field being modified. In other RPL specific replay attacks, the attacker primarily aims to introduce un-optimized or non-existent paths in the network by simply replaying the previously eavesdropped DIO packet after a certain period of time. The copycat attacker focuses on the combination of replay and interference method, unlike that of DIS flooding attack where fresh DIS packets are used to start a flood of DIO messages in the network. Although other routing attacks may also target QoS of IoT applications, the major difference lies in the way attacks are launched.}

\begin{figure}[!h]
	\centering
	\includegraphics[width=.8\textwidth]{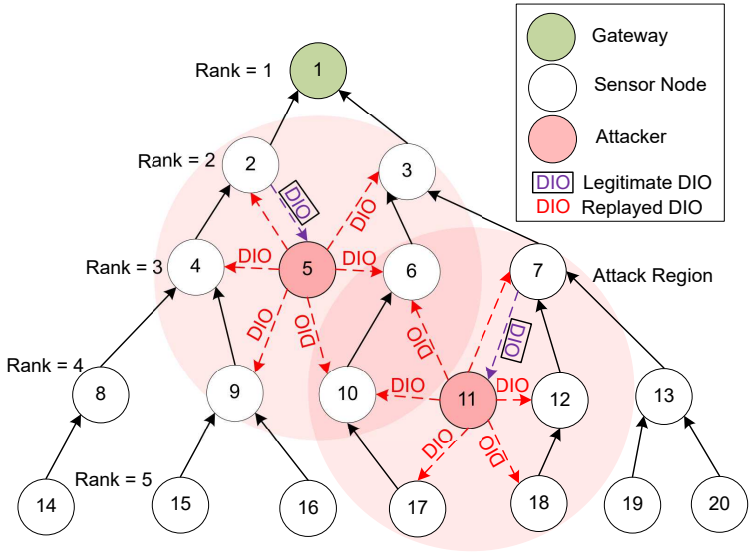}
	\caption{Illustration of copycat attack}
	\label{Fig:CopycatIllustration}
\end{figure}

\textcolor{black}{The standard RPL specification states that the link quality (e.g., Expected Transmission Count) must be computed before adding a new node in the candidate parent set.} Upon receiving the replayed DODAG Information Object (DIO) messages, a probing mechanism is initiated to assess the link quality. In this case, the probing fails because the replayed source is not in the node's communication range; hence the path is assumed to be bad and consequently discarded \cite{wallgren2013routing}. Thus the neighbor attack is ineffective if nodes are using ETXOF or MRHOF. \textcolor{black}{Also, a copycat attack with a fixed time interval keeps node busy continuously and, consequently, degrades the network's performance. It is to be noted that a copycat attack can also be performed with random intervals. However, the interval needs to be short in order to achieve maximum damage to the network. Also, adding a mechanism to compute random intervals very frequently will impose computational overhead to the attacker node, thereby decreasing the attacker node’s lifetime. Considering the fact that the attacker's primary target is to cause maximum damage to the network, it will simply choose a shorter interval (fixed value) and perform attack for a longer time.  In this study, we have considered the attack with fixed intervals.} \textcolor{black}{In spoofed copycat attack the attacker uses the source IP of one or more legitimate nodes (i.e., like Sybil attack); the attack will be ineffective if RPL is configured with MRHOF. Whereas in case the RPL is configured with OF0, then the attacker will succeed in persuading legitimate nodes that it is a potential parent. This is because the nodes do not check for neighbor reach-ability in case of OF0.}

\begin{figure}[!h]
	\centering
	\includegraphics[width=.8\textwidth]{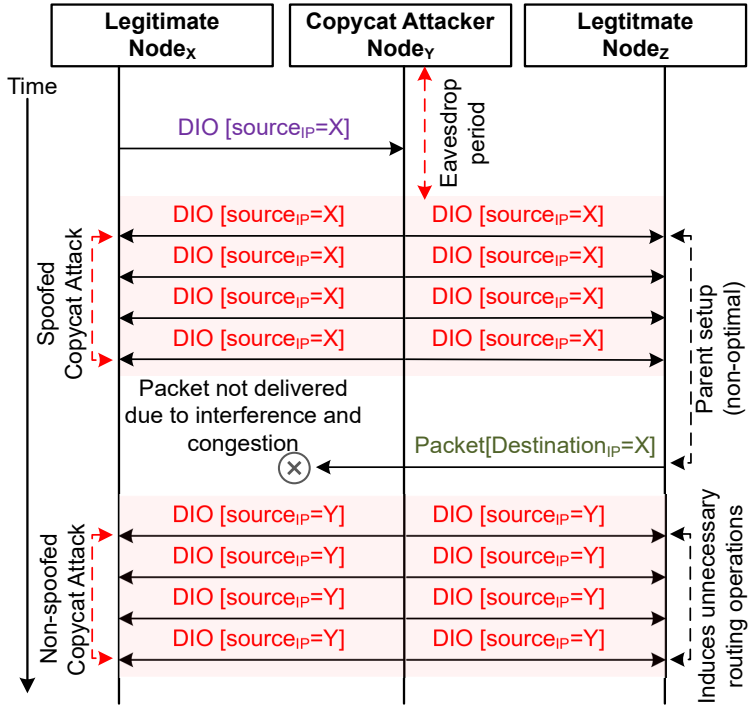}
	\caption{Attacker's approach for launching copycat attack}
	\label{Fig:CopycatFlow}
\end{figure}

The copycat attack can be illustrated using a simple example. Consider the network of 20 nodes depicted in Fig. \ref{Fig:CopycatIllustration}. As soon as the attacker nodes 5, 11 successfully eavesdrop the DIO sent from nodes 2 and 7, respectively, they start sending (multicast) the eavesdropped DIO to their neighbors. If an attacker is using the spoofed copycat attack approach, then nodes 6, 9, and 10 are forced to believe that node 2 is their in-range neighbor who actually is not. When the victim nodes 6, 9, and 10 try to send packets to the out of range neighbor (node 2), the packets are lost on the way without reaching the destination. In a scenario where an attacker is following a non-spoofed copycat attack approach, then the victim nodes are induced to believe that the sender of the DIO is legitimate, and hence they may try to add attacker node to their preferred parent list. \textcolor{black}{Moreover, when an eavesdropped packet is frequently replayed multiple times with a fixed interval, a heavy interference is introduced in the network region, i.e., an attacker's communication range. A Similar scenario is shown in Fig. \ref{Fig:CopycatIllustration}, where nodes 2, 3, 4, 6, 9, 10 are in range of attacker node 5, and nodes 4, 7, 10, 12, 17, 18 are in range of attacker node 11.} \textcolor{black}{The main reason for the significant drop in the performance of RPL based networks is the heavy congestion and interference introduced by the attacker in its attack region. Heavy congestion is induced due to the injection of a large number of DIO packets in the network, whereas interference is caused because the attacker's radio remains in transmitting state for most of the time. Both congestion and interference affect the packet forwarding behavior of legitimate nodes because of increased collisions, which consequently decreases PDR and increases AE2ED.} The non-spoofed and spoofed copycat attacks are illustrated in Fig. \ref{Fig:CopycatFlow}. The copycat attacker node is programmed so that it remains isolated (neither makes a parent nor becomes a parent) from the network while only performing a replay attack. In this way, an attacker can reduce its energy consumption rate for achieving a long-lasting attack.                

\textcolor{black}{Both the variants of a copycat attack can be performed using either insider or outsider attack strategy. In an insider attack strategy, the attacker compromises legitimate nodes to gain direct access to the network and reprograms them to launch the attack. Whereas, in an outsider attack strategy, the attacker does not directly access legitimate nodes; instead, it uses the physical medium (wireless communication channel) to launch the target network and launch the attack. In this study, we have assumed the insider attack strategy only. To launch a copycat attack using an outsider attack strategy, the attacker only needs to eavesdrop the packet transmissions by sensing the wireless communication channel. After capturing the DIO packet from a legitimate node, the attacker can launch a spoofed or non-spoofed copycat attack. The main reason that favors the possibility of launching a copycat attack using an outsider attack strategy is that the secure mode of RPL is not recommended due to energy constraints. Therefore, the attacker can quickly process unencrypted DIO packets.}

\section{Experimental Evaluation} \label{Experimental Evalutaion}
In this section, we discuss the simulation based impact analysis of copycat attacks. To evaluate the effects of copycat attacks on the RPL based network, various sets of experiments have been performed using the Cooja simulator, which is the most reliable and widely used network simulator and comes with the Contiki operating system. Contiki is a well known lightweight and publicly available operating system for constrained devices.     

\subsection{Objectives of the Experimental Study}
This paper aims to analyze and study the impact of the copycat attacks on the performance of the RPL based 6LoWPAN network by comparing the results of insecure RPL (under non-spoofed and spoofed copycat attack) with standard RPL protocol implementation (under no attack scenario). The primary objectives of the experimental study are:

\begin{enumerate}
	\item To analyze the performance of standard RPL implementation under no attack scenario.
	\item To show the effects of the spoofed and non-spoofed copycat attack on the RPL protocol.
	\item To investigate the factors that degrade the constrained nodes' performance, i.e., lifetime under copycat attacks.   
\end{enumerate}

\subsection{Experimental Setup}
Cooja has a hardware simulator named MSPsim that emulates the exact binary code of real sensor devices in order to achieve realistic simulation. In this paper, Zolertia 1 (Z1) platform (MSP430 architecture based ultra-low-power micro-controller board), which has the IEEE 802.15.4 compliant CC2420 radio transceiver operating at 2.4GHz, is used. The hardware specifications of the Z1 mote are shown in Table \ref{table3}. Table \ref{table4} presents the simulation parameters considered in the experiments. The Contiki operating system is modified to implement the copycat attack on attacker nodes. Specifically, an attacker node is programmed to eavesdrop and capture DIO messages from any legitimate node, and then replay the captured message with fixed replay interval. A network scenario containing one gateway node and 30 sensor nodes that are placed randomly on a grid of 200m $ \times $ 200m is considered. Each sensor sends a data packet of 30 bytes after every interval of 60 seconds. In order to perform fair experiments, the attacker nodes are programmed to activate after 90 seconds start of the network. In this way, the attack begins after the network is established and becomes stable. In order to simulate a realistic scenario, the Multipath Ray-Tracer Medium (MRM) radio model is used in all the experiments. \textcolor{black}{The simulation and MRM radio model parameters have been adopted from Perazzo \textit{et al.} \cite{Perazzo2017, vallatiemail} because the authors proposed a replay based attack that targets RPL, and therefore a similar topology will help assess the impact of the attack in a correct way.} We assume that the network is static, and the attacker has compromised some legitimate internal nodes and reprogrammed them to perform a copycat attack (insider attack scenario).      

\begin{table}[!h]
	\centering
	\caption{Hardware specification of Z1 mote \cite{advancare2010zolertia}}
	\begin{tabular}{|l|l|}
		\hline
		\textbf{Parameter} & \textbf{Value}\\ \hline
		Current consumption in CPU mode & 426$ \mu $A at 3V \\ \hline
		Current consumption in LPM mode & 20$ \mu $A  at 3V \\ \hline
		Current consumption in TX mode & 17.4mA at 3V\\ \hline
		Current consumption in RX mode& 18.8mA at 3V\\ \hline
		RAM size & 8KB \\ \hline
		ROM size & 92KB \\ \hline

	\end{tabular} 
	\label{table3}
\end{table}

\begin{table}[!h]
	\centering
	\caption{Simulation parameters}
	\begin{tabular}{|l|p{4.5cm}|}
		\hline
		\textbf{Parameter} & \textbf{Values}\\ \hline
		Radio model & Multipath Ray-Tracer Medium (MRM) \\ \hline
		Simulation area & 200m $ \times $ 200m\\ \hline
		Simulation time & 1800 seconds \\ \hline
		Objective function & Minimum Rank with Hysteresis Objective Function(MRHOF) \\ \hline
		Number of attacker nodes & 5 \\ \hline
		Number of gateway nodes & 1 \\ \hline
		Number of sensor nodes & 30 \\ \hline
		DIO minimum interval & 4 seconds \\ \hline
		DIO maximum interval & 17.5 minutes \\ \hline
		Replay interval & 1, 2, 3, 4 seconds \\ \hline
		Data packet size & 30 bytes \\\hline
		Data packet sending interval & 60 seconds \\ \hline
		Transmission power & 0dBm \\\hline
		
	\end{tabular} \label{table4}
\end{table}

\subsection{Performance Indicators}
In order to analyze the impact of the copycat attacks on the RPL based network, three prominent evaluation metrics are chosen. These metrics are Packet Delivery Ratio (PDR), Average End-to-End Delay (AE2ED), and Average Power Consumption (APC), which are defined as, 

\begin{enumerate}
	\item \textit{PDR}: It is the ratio between the total number of data packets received by the gateway node to the total data packets sent by the sensor nodes, including re-transmitted data packets in time interval $ T $. PDR is calculated as Eq. \ref{Eq-PDR}.
	
	\begin{equation}
		PDR = \frac{P_{received}}{\sum_{i=1}^{N} P_{sent_i}}
		\label{Eq-PDR}
	\end{equation} 
	where $ P_{received} $ represents the total number of data packets received at gateway node, and $ P_{sent_i} $ represents the total data packets sent from non-root node $ i $.
	
	\item \textit{AE2ED}: AE2ED is defined as the average amount of time taken by all the data packets sent from each sensor node, to be successfully delivered to the gateway node while neglecting all lost and dropped packets. AE2ED is calculated as Eq. \ref{Eq-AE2ED}.
	\begin{equation}
		AE2ED = \frac{\sum_{i=1}^{N} P_{received_i}}{P_{N}}
		\label{Eq-AE2ED} 
	\end{equation}
	where $ P_{received_{i}} $, $ P_{N} $ represent time delay of data packet $ i $ and total number of received packets, respectively. 
	\item \textit{APC}: It represents network wide power usage by the nodes. APC is based on two terms namely energy and power. Where energy usage refers to the energy consumed by the node's micro-controller unit (MCU) in transmission (TX), receiving (RX), low  power mode (LPM) (while MCU is in \textit{idle} mode but radio is \textit{OFF}) and CPU time (while MCU is \textit{ON} and the radio is \textit{OFF}). Eqs. \ref{energy}, \ref{power} represent energy and power, respectively.  
	
	\begin{equation}
		%	\begin{split}
		Energy(mJ) =
		(MCU_{TX} + MCU_{RX} + MCU_{CPU} + MCU_{LPM})
		\label{energy}
		%	\end{split}
	\end{equation}
	\begin{equation}
		Power(mW) = \frac{Energy}{Tos}
		\label{power}
	\end{equation}
	where $ Tos $ represents total operating time in seconds.
\end{enumerate} 

\subsection{Attack Implementation in Contiki}
The copycat attack is implemented by making modifications in Contiki source files. To add replay attack behavior \textit{rpl-timers.c} and \textit{rpl-icmp6.c} files have been modified. \textit{The handle\_dio\_timer()} and \textit{new\_dio\_interval()} methods of \textit{rpl-timers.c} are altered such that the attacker node replays eavesdropped DIO many times in fixed replay interval rather than replying on RPL's DIO Trickle timer. Similarly, the \textit{dio\_output()} method of \textit{rpl-icmp6.c} has been modified to replay eavesdropped DIO on every timer expire event (replay interval) while \textit{dio\_input()} is updated to only process and store the first DIO received by the attacker node. In addition, the functions required for other routing operations like topology management are disabled in order to save the attacker node's resources. To implement a spoofed copycat attack, \textit{uip\_icmp6\_send.c} has been modified to incorporate a mechanism that replaces the source IP address of the attacker node with a legitimate DIO sender's IP address in the ICMPv6 packet.             

\subsection{Experimental Results}
For each scenario, 10 independent replications with different seeds were run in order to obtain statistically valid results. The mean values of the obtained results with its errors at 95\% confidence interval have been reported to avoid biased observations. 

\begin{figure}[!h]
	\centering
	\includegraphics[width = .8\textwidth]{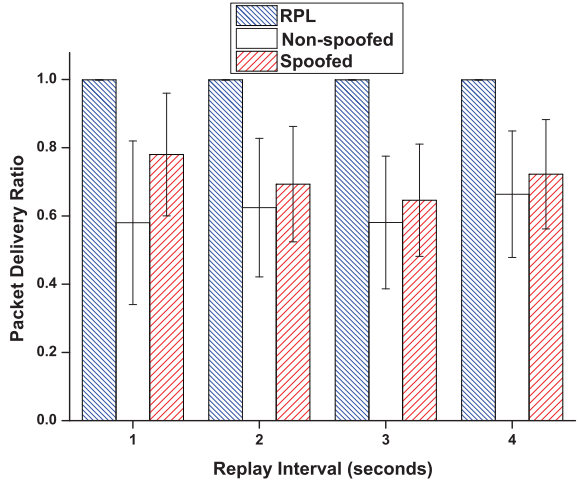}
	\caption{PDR values obtained in different scenarios}
	\label{PDR}
\end{figure}  

\subsubsection{Impact on Packet Delivery Ratio}\label{Impact on PDR}
Delivery of application data from sensor nodes to gateway is one of the major requirements of IoT applications. Thus, PDR analysis is an essential criterion in the performance evaluation of networks. Fig. \ref{PDR} shows the PDR values obtained with different replay intervals, i.e., $ 1, 2, 3,$ and $4 $ seconds. In order to perform a correct comparison, the results of the reference model (unattacked RPL) are also presented along with the attacked scenario (spoofed and non-spoofed). In the case of the reference model (RPL), it must be noted that the replay interval plays no role. It can be observed that spoofed and non-spoofed copycat attacks significantly bring down the networks PDR compared to the reference model's PDR. Moreover, by observing the values of PDR's of all three scenarios, we get a better insight into which attack has a higher impact on RPL protocol. The reference model shows an average PDR value of approximately $ 1 $, and in case of a non-spoofed attack, the PDR reduces to value between $ 0.58 $ to $ 0.66 $, whereas with a spoofed attack, the PDR reduces to value between $ 0.64 $ to $ 0.77 $. The main observation we made from the simulation experiments is that there is no linear relation between increasing replay interval values and PDR values. For all four replay interval, the reduction in PDR values is observed. \textcolor{black}{It can be seen that the non-spoofed copycat attack makes more damage to the network than the spoofed copycat attack. This is because, in case of a non-spoofed attack, the victim node receives DIO (with a non-spoofed source IP address) message from an unresponsive attacker multiple times, i.e., an attacker does not respond to victim's DAO messages. This forces the victim node to perform unnecessary routing management related operations on every illegitimate DIO reception, which limits its data packet forwarding behavior. While, in case of a spoofed attack, upon receiving a replayed DIO, a victim may add the distant node as its preferred parent assuming the DIO sender leads to an optimal route to the gateway. Thus, when the victim sends its data packets to a distant parent, the probability of packet getting lost increases due to the non-optimal route, congestion, and interference, and this causes a reduction in PDR. However, the redundant routing related operations are not invoked because DIO is received with the same source IP address causing DIO suppression in case of short replay intervals. This is the main reason to explain why the packet-forwarding behavior of victim nodes under non-spoofed copycat attacks is more disrupted as compared to spoofed copycat attacks.} Such a reduction in PDR values is not acceptable for critical IoT applications like healthcare. Hence copycat attacks must be addressed carefully for the smooth operation of such applications.

\begin{figure}[!h]
	\centering
	\includegraphics[width = .8\textwidth]{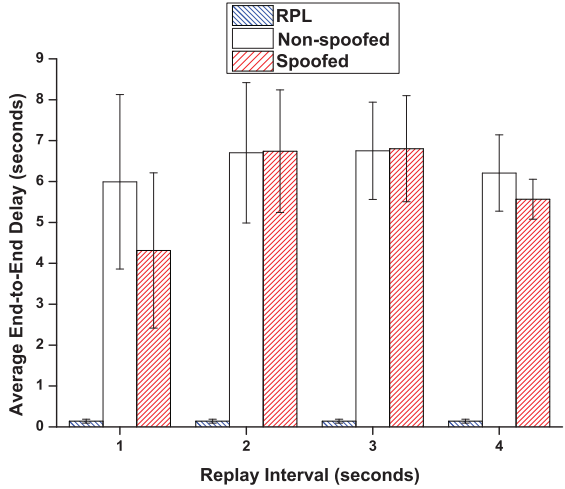}
	\caption{AE2ED values obtained in different scenarios}
	\label{AE2ED}
\end{figure}

\subsubsection{Impact on Average End-to-End Delay}
In Fig. \ref{AE2ED}, the mean AE2ED values of reference model (RPL), non-spoofed, and spoofed copycat attack scenarios are shown. The copycat attack significantly affects the AE2ED of the network by approximately $ 44 $ times than that of the reference model's AE2ED. As can be seen, the AE2ED value obtained in the reference model scenario is roughly $ 0.14 $ seconds. Whereas, in the case of a non-spoofed copycat attack, the value of AE2ED is between $ 5.99 $ to $ 6.75 $ seconds. With a spoofed copycat attack scenario, the mean value of the AE2ED value is between $ 4.31 $ to $ 6.80 $ seconds. The non-spoofed copycat attack scenario shows the lowest AE2ED value with a $ 1 $ second replay interval while the highest value obtained is in case of a $ 3 $ second replay interval. Like the non-spoofed attack scenario, the lowest and highest values obtained in the spoofed attack scenario are also $ 1 $ and $ 3 $ seconds, respectively. As observed from PDR analysis performed in the previous subsection \ref{Impact on PDR}, in this case, also, a non-linear relationship between increasing values of replay interval and AE2ED is observed from the experiments. \textcolor{black}{The increase in the value of AE2ED can be credited to two reasons: (1) congestion and interference evoked by the copycat attack which affects the forwarding nodes in the attack region; (2) creation of non-optimal routes due to replay of old routing information which leads to longer routing path for data packets.}

\subsubsection{Impact on Average Power Consumption}
\begin{figure}[!h]
	\centering
	\includegraphics[width = .8\textwidth]{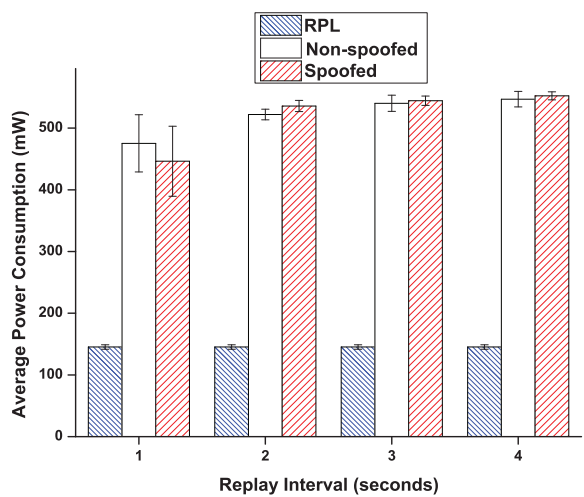}
	\caption{APC values obtained in different scenarios}
	\label{APC}
\end{figure}

The lifetime of a 6LoWPAN network is highly dependent on the power consumption of nodes in the network. The major problem arises when constrained nodes start consuming more power than normal consumption. This situation reduces network lifetime and degrades network performance, consequently affecting the services of target IoT applications. In this regard, we focus on analyzing the APC of the network. It can be observed from Fig. \ref{APC} that APC of the reference model is lesser than attack scenarios, which shows that the copycat attack significantly degrades the network lifetime. The copycat attack forces the node's radio to remain active for a longer time than normal, thereby increasing the power consumption. The copycat attack significantly affects the APC of the network by approximately $ 3.5 $ times than that of the reference model's APC. The reference model scenario shows the APC of approximately $ 145mW $ whereas non-spoofed and spoofed copycat attack scenario show APC between $ 475mW $ to $ 576mW $, and $ 446mW $ to $ 552mW $, respectively. It can be observed that APC under attack scenario follows a non-linear relationship with replay interval values. \textcolor{black}{The reason for this is that the nodes do not reset their trickle timer upon reception of every DIO that contains consistent routing information within the node's current DIO interval. Thus, varying replay interval does not give any extra benefit to the copycat attacker.} In every attack scenario, the obtained APC values are almost similar. The only small difference is observed in the case of 1 second interval. However, by observing the error bars, it can be said that in some experiments, the APC values obtained with 1 second interval are similar to those which obtained with 2, 3, and 4 second interval. Both spoofed and non-spoofed copycat attacks have an almost similar impact on the network's APC at a common replay interval. This is because of the similarity in implementing both non-spoofed and spoofed copycat attacks where an attacker replays the captured packets at a fixed interval. Therefore, the duration of time a victim node spends in CPU, LPU, TX, and RX mode are almost the same in both the attack scenarios (at common replay interval). Therefore the energy consumed by the victim node is almost the same under both the attack scenarios. To advocate our previous statement, we performed a more extensive study on several non-attacker nodes' power consumption. For the sake of simplicity, we discuss the power consumption of node 2 only in terms of CPU, LPM, TX, and RX considering both types of attacks at various replay intervals. It is to be noted that the graphs presented for analyzing the power consumption of node 2 are from single execution. Hence the mean values and errors at 95\% confidence interval are not shown.

\begin{figure}[!h]
	\centering
	\includegraphics[width = .8\textwidth]{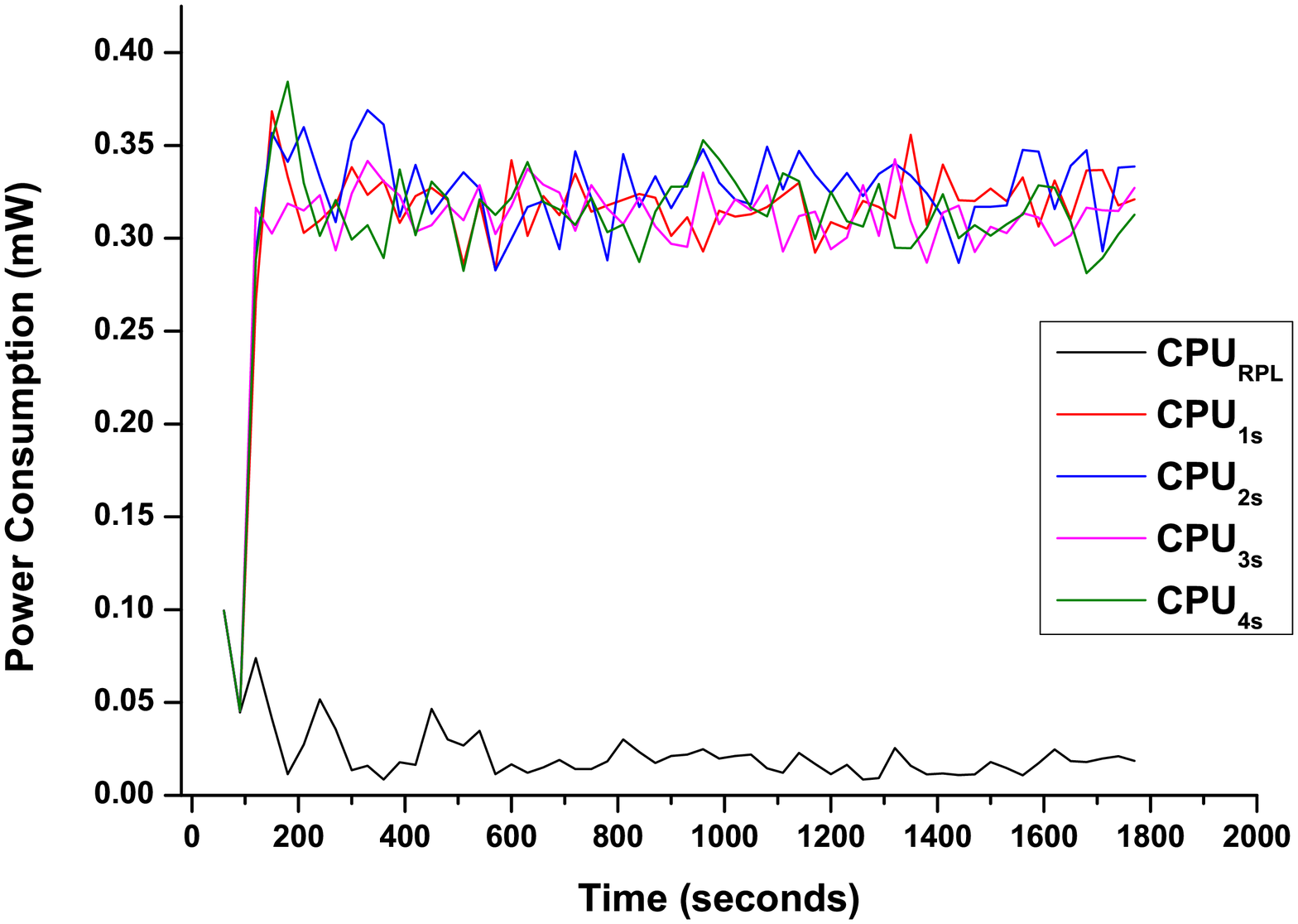}
	\caption{Power consumption by node 2 in CPU mode (Non-spoofed copycat attack)}
	\label{Node_2_CPU_Profile_S}
\end{figure}

\begin{figure}[!h]
	\centering
	\includegraphics[width = .8\textwidth]{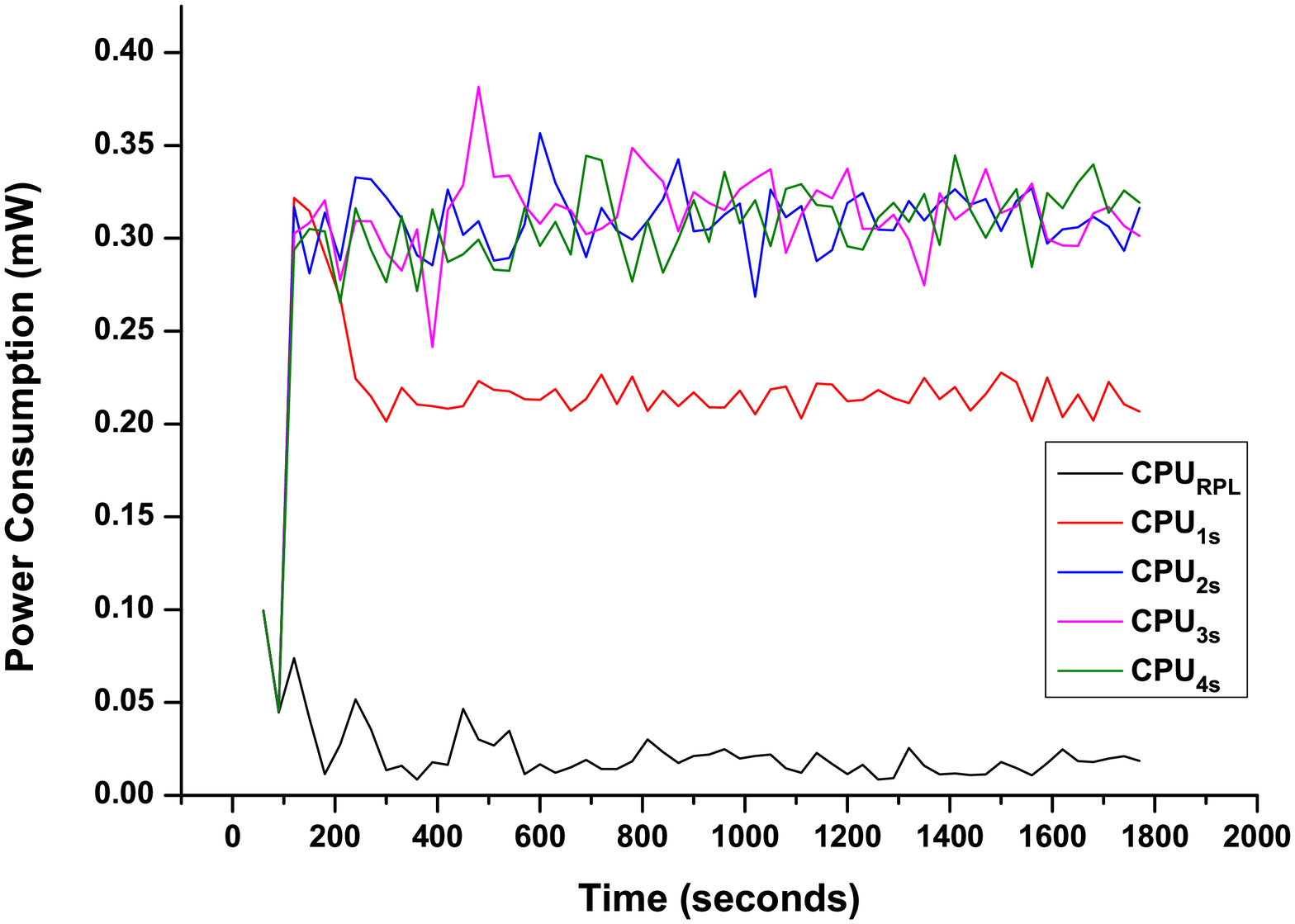}
	\caption{Power consumption by node 2 in CPU mode (Spoofed copycat attack)}
	\label{Node_2_CPU_Profile_NS}
\end{figure}
\begin{figure}[!h]
	\centering
	\includegraphics[width = .8\textwidth]{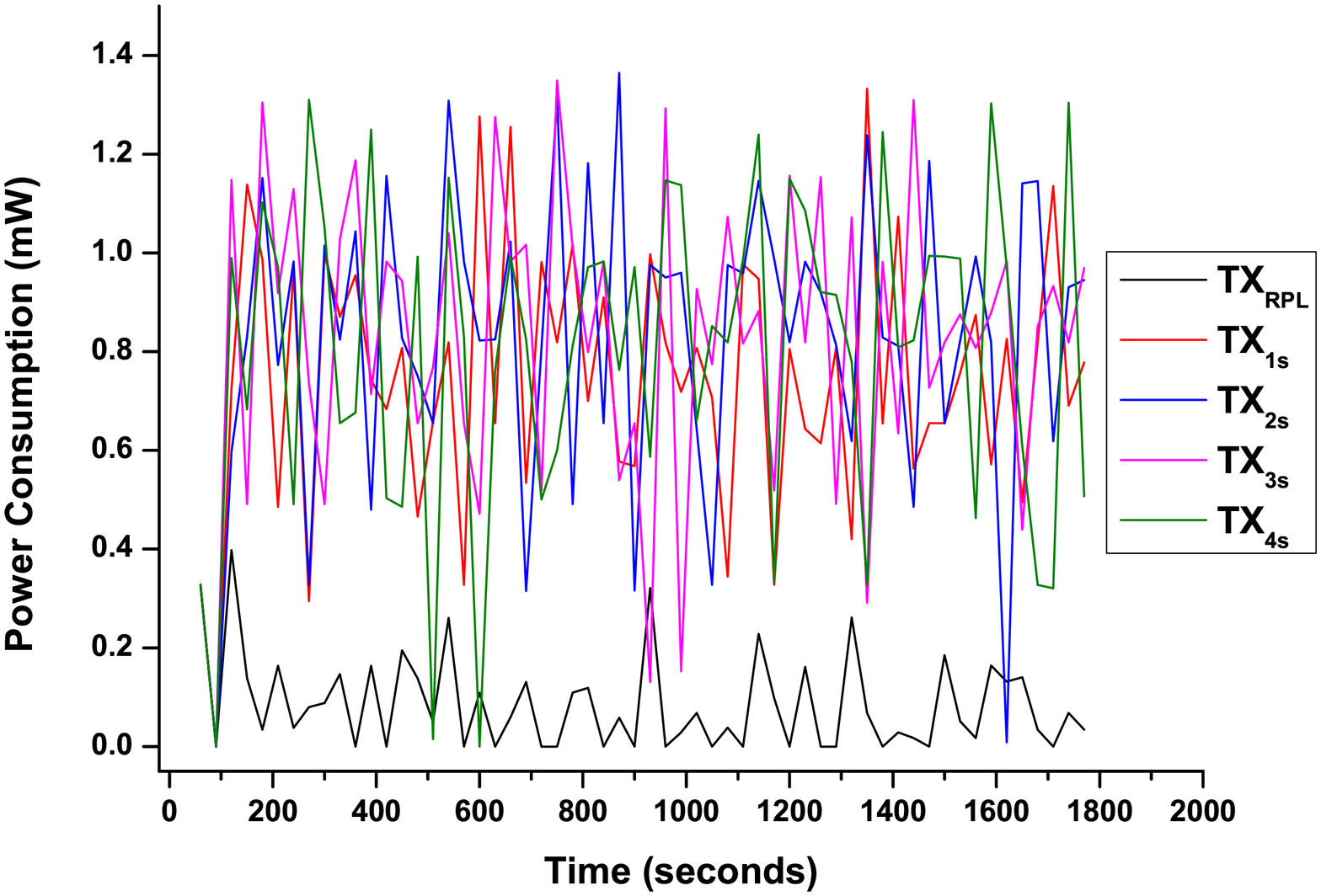}
	\caption{Power consumption by node 2 in TX mode (Non-spoofed copycat attack)}
	\label{Node_2_TX_Profile_S}
\end{figure}

\begin{figure}[!h]
	\centering
	\includegraphics[width = .8\textwidth]{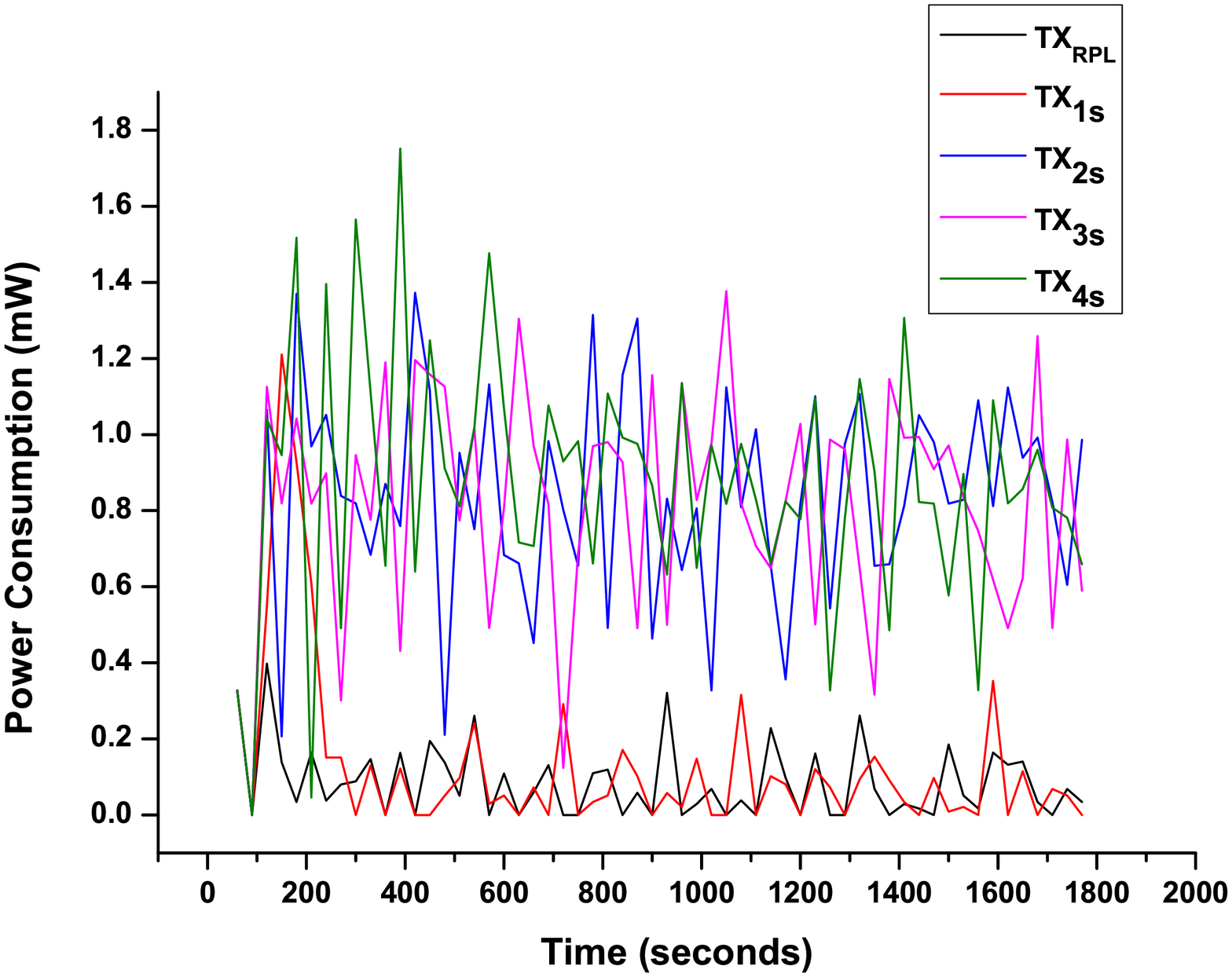}
	\caption{Power consumption by node 2 in TX mode (Spoofed copycat attack)}
	\label{Node_2_TX_Profile_NS}
\end{figure}

\begin{figure}[!h]
	\centering
	\includegraphics[width = .8\textwidth]{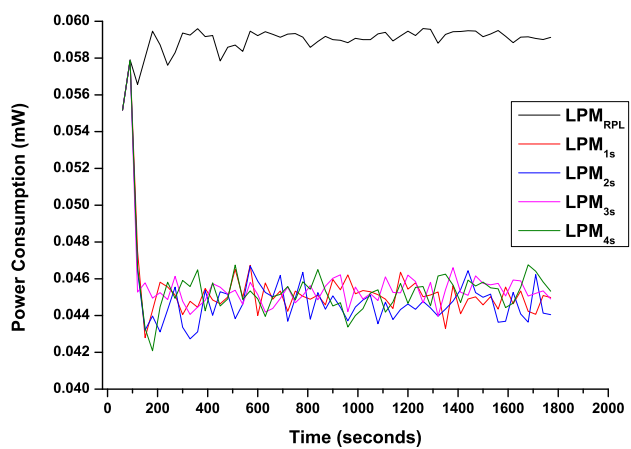}
	\caption{Power consumption by node 2 in LPM mode (Non-spoofed copycat attack)}
	\label{Node_2_LPM_Profile_S}
\end{figure}

\begin{figure}[!h]
	\centering
	\includegraphics[width = .8\textwidth]{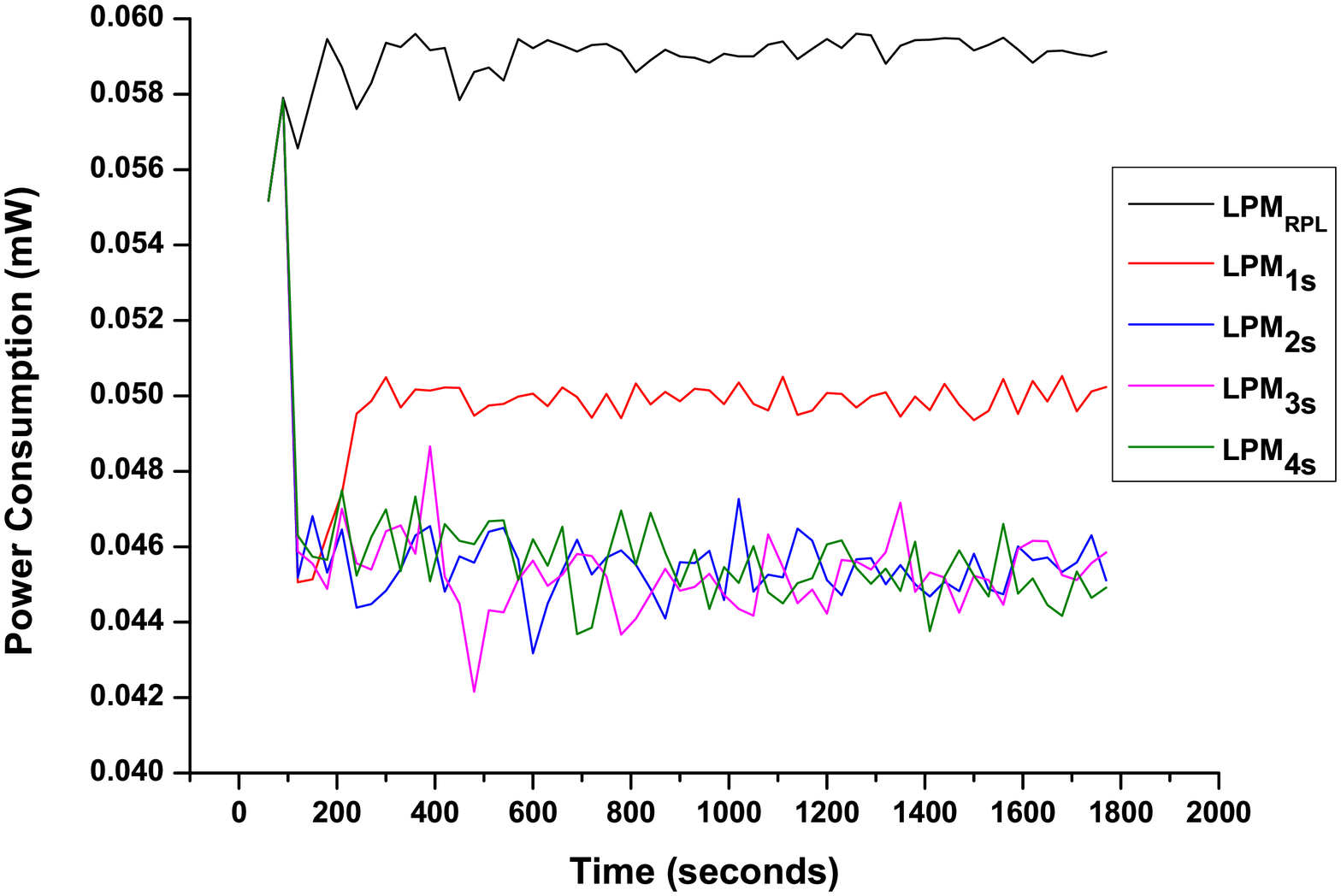}
	\caption{Power consumption by node 2 in LPM mode (Spoofed copycat attack)}
	\label{Node_2_LPM_Profile_NS}
\end{figure}

\begin{figure}[!h]
	\centering
	\includegraphics[width = .8\textwidth]{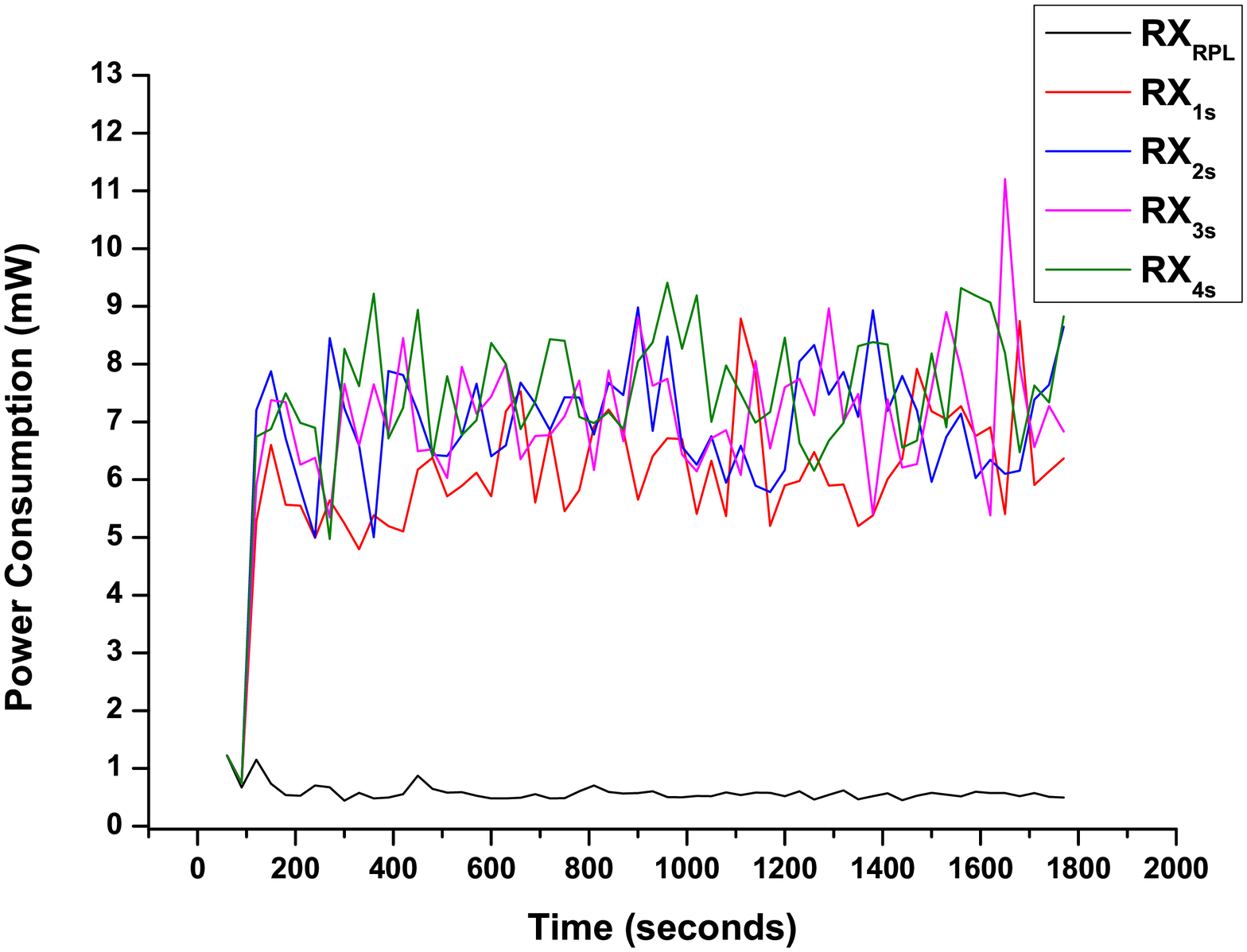}
	\caption{Power consumption by node 2 in RX mode (Non-spoofed copycat attack)}
	\label{Node_2_RX_Profile_S}
\end{figure}

\begin{figure}[!h]
	\centering
	\includegraphics[width = .8\textwidth]{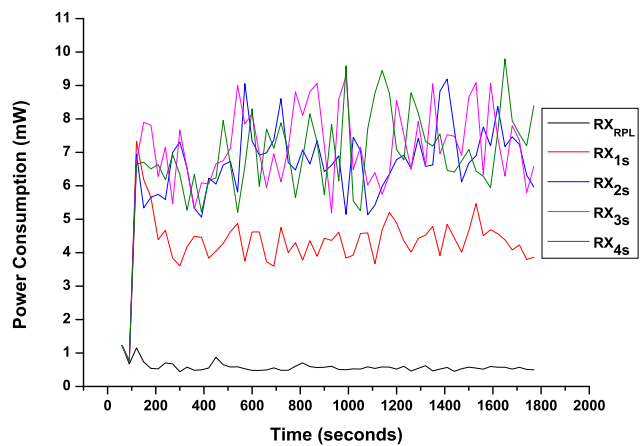}
	\caption{Power consumption by node 2 in RX mode (Spoofed copycat attack)}
	\label{Node_2_RX_Profile_NS}
\end{figure}

Fig. \ref{Node_2_CPU_Profile_S} and  Fig. \ref{Node_2_CPU_Profile_NS} shows the time-wise power consumption by node 2 in CPU mode under non-spoofed and spoofed copycat attack scenarios, respectively. It can be observed that the CPU power consumption of node 2 under non-spoofed and spoofed attack scenarios follow a similar trend over $ 2, 3, $, and $ 4 $ seconds replay interval. Whereas over $ 1 $ second replay interval, the power consumption under non-spoofed and spoofed attacks differs slightly. The LPM power consumption by node 2 also shows a similar pattern for replay interval of $ 2, 3, $, and $ 4 $ seconds, whereas differ in $ 1 $ second replay interval under both the types of copycat attacks as observed in Fig. \ref{Node_2_LPM_Profile_S} and Fig. \ref{Node_2_LPM_Profile_NS}. The power consumption in terms of TX mode for the non-spoofed and spoofed copycat attack is shown in Fig. \ref{Node_2_TX_Profile_S} and Fig. \ref{Node_2_TX_Profile_NS}, respectively. The TX power profile of node 2 under both attacks shows a variation in transmission power consumption with increasing time. It is also observed that the trend is very much similar for the replay interval of 2, 3, and 4 seconds, but different on 1 second replay interval. RX mode power consumption is depicted in Fig. \ref{Node_2_RX_Profile_S} and Fig. \ref{Node_2_RX_Profile_NS}. We observe an almost similar trend in RX power consumption along increasing time for both the types of copycat attacks. This is because the total RX time in a copycat attack scenario depends mainly on the replay interval of replayed packets. 

The reason for different results observed in the case of 1 second interval is that node 2 receives frequently replayed packets from the non-spoofed attacker, i.e., an attacker that does not hide its original identity or IP. Thus it is more likely that the attacker may become a potential candidate of preferred parent list. Therefore, a legitimate node is forced to perform unnecessary RPL related routing management operations upon receiving packets from a non-spoofed attacker. The same can be observed in Fig. \ref{Node_2_CPU_Profile_S}, \ref{Node_2_LPM_Profile_S}, and \ref{Node_2_TX_Profile_S}. Contrary to this, a spoofed copycat attacker replay packets after modifying the captured packet's source IP field, thus upon receiving a frequently replayed packet from a spoofed copycat attacker, a legitimate node increases its DIO redundancy counter and does not perform many RPL related routing management operations. Hence, a legitimate node does not spend more time in CPU, TX mode, and can be verified from the results shown in Fig. \ref{Node_2_CPU_Profile_NS}, \ref{Node_2_LPM_Profile_NS}, and \ref{Node_2_TX_Profile_NS}. 

\textcolor{black}{It is to be noted that the attacker’s success rate or performance depends on the placement of malicious nodes. The attack impact on the network is directly proportional to the number of legitimate nodes in attacker's range. It is observed that there is no significant difference between the attacker's impact on the network in case of a spoofed and non-spoofed copycat attack. The major difference is in the way both attacks are launched. In the case of a non-spoofed copycat attack, the detection is possible by merely analyzing the number of packets received from neighbors. In contrast, the detection of a spoofed copycat attack is typical because, in this case, the attacker spoofs its IP address and resembles itself as a legitimate sender. Although the spoofed copycat attacker may fail when MRHOF is enabled, it is still a significant threat considering that not all networks may be configured with MRHOF.  This study aims to analyze the impact of both attack variants so that suitable solutions can be developed in the future.} 

\section{\textcolor{black}{Possible Solutions for Mitigation of Copycat Attack}}\label{PossibleSol}
\textcolor{black}{To mitigate copycat attacks, a simple solution is to analyze the distribution of DIO packets received from neighbors. Copycat attacks involve frequent replaying of DIO packets. In RPL, the rate of transmission of control messages is controlled by trickle timer to minimize energy consumption, but in case of attack, the attacker transmits a large number of DIO packets in the network. Outlier detection can be used to find the neighbors showing abnormal behavior \cite{barnett1974outliers,verma2019addressing,vermacosec}. In this case, abnormal behavior can be thought of as the neighbor that sent a large number of DIO packets compared to other neighbors. Outlier detection involves the detection and removal of outliers from the data. Outlier detection problem can be mapped to the intrusion detection problem of RPL based 6LoPWANs \cite{kumar2016anomaly,jabez2015intrusion}. An outlier can be assumed as a node with abnormal behavior (i.e., malicious node), which needs to be identified and eliminated to achieve better network performance. Different methods can be used to build an outlier detection model for RPL. Some of these techniques include Interquartile Range, Kalman filter (statistics and control theory), and Entropy (Information theory) \cite{wang2018robust, domingues2018comparative,zhi2018gini}. Another possible solution is to use IPv6 over the Time Slotted Channel Hopping (TSCH) mode of IEEE 802.15.4e (6TiSCH) \cite{dujovne20146tisch}. TSCH technique secures the network from jamming attacks, and therefore it can help mitigate the effect of the attacker that is transmitting DIO packets at a high rate \cite{zorbas2018r}. } 

%\newpage
\section{Conclusions and Future Scope}\label{Conclusions and Future Scope}

This paper has investigated the copycat attack, a type of DoS attack which forces legitimate nodes to believe in false routing information while introducing congestion and interference in the RPL based network. This situation leads to the creation of the non-optimal routes, which severely degrades the network performance. An attacker can perform the copycat attack by either insider or outsider attack strategy. Moreover, an attacker does not require any specialized hardware for launching the copycat attack from IoT nodes. We have shown that this attack severely degrades the routing and application performance using an in-depth experimental study. In the future, we aim to study the effects of copycat attack in dynamic networks and implement a security mechanism to defend such attacks.

\section*{Acknowledgments}
This research is supported by the Ministry of Human Resource Development (MHRD), Government of India.

\section*{References}

\end{document}